\documentclass[preprintnumbers,amsmath,amssymb,aps,prl]{revtex4}
\usepackage{amsmath}
\usepackage[mathcal]{eucal}
\usepackage{latexsym}
\usepackage{amssymb}
\usepackage{color}
\definecolor{Red}{rgb}{1.,0.,0.}
\definecolor{Blue}{rgb}{0.,0.,1.}

\newcommand*{\eps}{{\rlap{\lower2ex\hbox{$\,\,\tilde{}$}}{\epsilon_{ijk}}}}
\newcommand*{\EPS}{{\rlap{\lower2ex\hbox{$\,\,\tilde{}$}}{\epsilon_{i'j'k'}}}}
\newcommand*{\lmq}{{\rlap{\lower2ex\hbox{$\,\,\tilde{}$}}{\epsilon_{lmq}}}}
\newcommand*{\jmq}{{\rlap{\lower2ex\hbox{$\,\,\tilde{}$}}{\epsilon_{jmq}}}}
\newcommand*{\jql}{{\rlap{\lower2ex\hbox{$\,\,\tilde{}$}}{\epsilon_{jql}}}}
\newcommand*{\jlm}{{\rlap{\lower2ex\hbox{$\,\,\tilde{}$}}{\epsilon_{jlm}}}}
\newcommand*{\imq}{{\rlap{\lower2ex\hbox{$\,\,\tilde{}$}}{\epsilon_{imq}}}}
\newcommand*{\iql}{{\rlap{\lower2ex\hbox{$\,\,\tilde{}$}}{\epsilon_{iql}}}}
\newcommand*{\ilm}{{\rlap{\lower2ex\hbox{$\,\,\tilde{}$}}{\epsilon_{ilm}}}}
\newcommand*{\lmn}{{\rlap{\lower2ex\hbox{$\,\,\tilde{}$}}{\epsilon_{lmn}}}}
\newcommand*{\abc}{{\rlap{\lower2ex\hbox{$\,\,\tilde{}$}}{\epsilon_{abc}}}}
\newcommand*{\N}{{\rlap{\lower2ex\hbox{$\,\,\tilde{}$}}{N}}}
\newcommand{\tN}{{\rlap{\lower2ex\hbox{$\,\,\tilde{}$}}{N}}}
\newcommand*{\tM}{{\rlap{\lower2ex\hbox{$\,\,\tilde{}$}}{M}}}
\newcommand*{\I}{{\rlap{\lower2ex\hbox{$\,\,\tilde{}$}}{e_{i}^{a}}}}
\newcommand*{\J}{{\rlap{\lower2ex\hbox{$\,\,\tilde{}$}}{e_{j}^{a}}}}

\begin{document}
\title{A proposal for a reduced phase space for time symmetric, Lorentzian gravity}

\author{Eyo Ita}\email{ita@usna.edu}
\address{Department of Physics, US Naval Academy, Annapolis Maryland}
\input amssym.def
\input amssym.tex

%\address{$^{1}$\,\,Department of Physics, US Naval Academy,
%Annapolis, Maryland}
%\address{$^{2,3}$\,\,Department of Physics, National Cheng Kung University,
%Tainan 701, Taiwan}

%\ead{ita@usna.edu$^{1}$, cpsoo@mail.ncku.edu.tw$^{2}$,
%ccmaple014@gmail.com$^{3}$}

\bigskip

\begin{abstract}
In this paper we perform a full gauge-fixing of the phase space of four dimensional General Relativity (GR) of Lorentzian signature for the time symmetric case, using the CDJ variables.  In particular, the Gauss' law constraint in the chosen gauge meets the conditions of the Cauchy-Kovalevskaya theorem for first order, quasilinear PDEs.  This implies the existence of a unique analytic solution to the initial value constraints problem in some region of 3-space, featuring four free functions per spacetime point.  This result constitutes a step toward addressal of the reduced phase space problem of GR.
\end{abstract}
\maketitle

\section{I. Introduction}

The task of writing down a general solution to the Einstein field equations has been a main unsolved problem in physics since the introduction of four dimensional General Relativity (GR).  A natural question to ask is how closely one can come, in some suitable sense, to writing down such a solution.  In this paper, we will broach this topic from the perspective of the initial value constraints problem.  Given a 3+1 formulation of gravity with phase space variables $(q,p)$, where $q$ refers to the coordinate and $p$ the momentum, one has constraint equations and evolution equations.  The constraint equations $C(q,p)=0$ place constraints on the allowable initial data $q,p$, and the evolution equations prescribe the velocities $\dot{q},\dot{p}$ which govern the evolution of this data.  A GR solution would consist of solving the constraints, imposing one gauge-fixing (subsidiary) condition for each constraint, such that in the end one has four phase space degrees of freedom point.  As a matter of consistency, the constraints and subsidiary conditions must be preserved under time evolution.  This problem has been addressed in the linearized approximation to gravity in various approaches.  But the construction of solutions to the constraints in the full nonlinear theory, while imposing `good' gauge-fixing conditions, has thus far proven to be a challenging problem which remains unaddressed.\par 
\indent 
The standard approach to totally constrained systems due to Dirac \cite{DIR} relies on a canonical structure and the ability to define Poisson brackets and Dirac brackets, primary and secondary constraints, etc.  But in carrying out the remaining steps of the Dirac procedure, one would be eventually faced with the problem of constructing a reduced phase space for gravity, which in the full theory still remains an unsolved problem.\par 
\indent 
This paper will constitute an attempt to understand the physical phase space of full, nonlinearized gravity, and to probe deeper into its mathematical structure.  The Ashtekar variables provide a good starting point in that the initial value constraints and evolution equations can be written as low order polynomials.  The intractability of the constraint and gauge fixing equations leads one to try out a different set of variables $A,\Psi$, the CDJ variables attributed to Riccardo Capovilla, John Dell and Ted Jacobson.  The constraint equations in these variables are mostly algebraic and, as we will see, the evolution equations contain relatively few spatial derivatives acting on the Lagrange multiplier fields.  We will see that this simplifies the process of solving the constraints and gauge-fixing, which will provide accessibility to larger sectors of the reduced phase space of gravity.\par 
\indent 
Another problem is the issue of reality conditions.  The viewpoint presented in this paper is that the process of obtaining a reduced phase space and that of implementing reality conditions are two separate problems, whether one uses the Ashtekar variables or one uses $A,\Psi$.  Moreover, one can always attempt to obtain real GR solutions of Lorentzian signature by considering the time-symmetric case in either approach.  Then one still has the initial value constraints/gauge fixing problem to deal with, which while still complicated in the Ashtekar variables becomes vastly simplified for $A,\Psi$.  So in this paper we will focus on the aspects of solving GR strictly related to the reduced phase space problem, which is separate from the issue of reality conditions.  We will show under fairly general conditions that there exists a class of solutions specified by four free functions per spacetime point.
\subsection{Organization of this paper}
This paper is structured as follows.  Section II starts off with the Ashtekar variables, and proceeds directly to the CDJ $A,\Psi$ variables, rewriting the constraints and the evolution equations.  We reinterpret the problem in terms of transformations of the new phase space $A,\Psi$.  Section III is the gauge-fixing section, where we provide physical motivations for the desired gauge-fixing choice, and demonstrate that the gauge is indeed accessible.  In Section IV we solve the initial value constraints in the chosen gauge.  In particular, we propose a solution to the Gauss' law constraint.  This is made possible due to a particular aspect of the gauge-fixing choice, known (from Yang--Mills theory) as the axial gauge.  On account of this, the Gauss' constraint reduces to a particular combination of components such that the Cauchy--Kovalevskaya theorem for first order quasilinear systems of PDEs becomes applicable.  The proposed solution features four freely specifiable, real-analytic functions corresponding to the physical degrees of freedom of GR in this gauge.  Section V is a summary, delineating the main results.\par 
\indent

\section{II. The Ashtekar variables}

Let $M$ be a four dimensional spacetime manifold of topology $M=\Sigma\times{R}$, for some 3 dimensional spatial manifold $\Sigma$ of a given topology.  Then the Ashtekar action for gravity with cosmological constant $\Lambda$ can be written as \cite{ASH1}, \cite{ASH2}, \cite{ASH3}
\begin{eqnarray}
\label{STARTING}
I=\int{dt}\int_{\Sigma}d^3x\biggl[\widetilde{\sigma}^i_a\dot{A}^a_i+A^a_0D_i\widetilde{\sigma}^i_a-\epsilon_{ijk}\widetilde{\sigma}^i_aB^j_aN^k
-iN\sqrt{\hbox{det}\widetilde{\sigma}}\bigl(\Lambda+B^i_d(\widetilde{\sigma}^{-1})^a_i\bigr)\biggr].
\end{eqnarray}
\noindent 
The Ashtekar phase space variables are a gauge connection $A^a_i$ and densitized triad $\widetilde{\sigma}^i_a$, given by\footnote{For index labelling conventions, beginning Latin letters $a,b,c,\dots$ denotes internal $SO(3)$ indices, while those from the middle $i,j,k,\dots$ denote spatial indices.  Each index set takes values $1-3$.} 
\begin{eqnarray} 
\label{USING}
A^a_i=\Gamma^a_i+\beta{K}^a_i;~~\widetilde{\sigma}^i_a={1 \over 2}\epsilon^{ijk}\epsilon_{abc}e^b_je^c_k, 
\end{eqnarray}
\noindent 
where $\beta$ ($-i$ for the self-dual case) is the Immirzi parameter, $\Gamma^a_i$ is the connection compatible with the spatial triads $e^a_i$, and $K^a_i=e^a_jK^j_i$ is the triadic form of the extrinsic curvature of 3-space.  
The initial value constraints, the Gauss' law, vector and Hamiltonian constraints, are given by the Lagrange's equations for the temporal component of the spacetime connection $A^a_{\mu}$ and the shift vector and lapse function, the auxiliary fields $A^a_0$, $N^i$ and $N$ 
\begin{eqnarray}
\label{ASHVARIABLES}
D_i\widetilde{\sigma}^i_a=0;~~\epsilon_{ijk}\widetilde{\sigma}^j_aB^k_a=0;~~\epsilon_{ijk}\epsilon^{abc}\widetilde{\sigma}^i_a\widetilde{\sigma}^j_b\Bigl(B^k_c+{\Lambda \over 3}\widetilde{\sigma}^k_c\Bigr)=0.
\end{eqnarray} 
\noindent 
The constraints are the canonical manifestation of gauge symmetries in the theory and define a constraint surface, signifying that not all of the starting phase space is accessible to the system.  The equations of motion for the dynamical variables can be obtained by varying the Lagrangian (\ref{STARTING}) with respect to the densitized triad, yielding
\begin{eqnarray} 
\label{USING}
\dot{A}^a_i=D_iA^a_0+\epsilon_{ijk}B^j_aN^k-iN\sqrt{\hbox{det}\widetilde{\sigma}}B^n_d(\widetilde{\sigma}^{-1})^d_i(\widetilde{\sigma}^{-1})^a_n,
\end{eqnarray} 
\noindent 
and with respect to the connection, yielding
\begin{eqnarray} 
\label{THECONNECTION1}
\dot{\widetilde{\sigma}^i_a}=f_{abc}\widetilde{\sigma}^i_bA^c_0+\epsilon_{mjk}\epsilon^{jni}D_n(\widetilde{\sigma}^m_aN^k)
-i\epsilon^{ijk}D_j(N\sqrt{\hbox{det}\widetilde{\sigma}}(\widetilde{\sigma}^{-1})^a_k).
\end{eqnarray}
\noindent 
The evolution of the system from a given set of initial conditions must respect the constraints (\ref{ASHVARIABLES}) and is not unique, on account of the presence of gauge symmetries in the theory.  To find the physical degrees of freedom one must perform a gauge-fixing procedure corresponding to each constraint (\ref{ASHVARIABLES}), while solving the constraints within the chosen gauge.\footnote{The reduced phase dimension should be $9+9=18$ unconstrained phase space variables minus $7$ constraints minus $7$ gauge fixing conditions, or $4$ phase space degrees of freedom per point.}  Additionally, appropriate reality conditions must be implemented on the phase space in order to ensure that in the end one has GR corresponding to real-valued spacetime metrics.\par 
\indent 
The construction of a reduced phase space for gravity has been a long-standing open problem.  Let us outline some of the main difficulties along with how we plan to address them.\par 
\noindent 

\subsection{II.1. Addressing the difficulties: The CDJ variables}

(i) Reality conditions:  For $\beta=\pm{1}$ the reality conditions can be satisfied by choosing all variables to be real-valued, which corresponds to Euclidean signature GR.  For $\beta=\pm{i}$ one obtains Lorentzian signature GR which is in general complex, and the reality conditions are nontrivial to impose.  For nondegenerate triads, $K^a_i=0$ implies the condition of time symmetry $K_{ij}=0$, namely that the extrinsic curvature must be zero on the initial spatial slice $\Sigma_0$.  It is possible that $K_{ij}$ can become nonzero for $t\neq{0}$.  From (\ref{USING}) the connection $A^a_i$, which is real-valued at $t=0$, could then become complex for $t\neq{0}$.  Then the next concern would be whether the spatial 3-metric $h_{ij}$, which started out real at $t=0$, could also become complex under this evolution.  However, the metric is guaranteed to remain real-valued, since the reality conditions in the Ashtekar variables must be preserved under time evolution \cite{ASH1}.  So while the time symmetric case is a special case, it does nevertheless provide a nontrivial sector thereof of real-valued Lorentzian spacetime metrics solving the Einstein equations.  The results of this paper, for Lorentzian signature solutions, will be confined to the time symmetric case.\par 
\noindent 
(ii) Constraints: The initial value constraints (\ref{ASHVARIABLES}), while low order polynomials in the Ashtekar variables, are still nontrivial to solve.  There is a certain substitution due to authors  Riccardo Capovilla, John Dell and Ted Jacobson, which we will refer to as the `CDJ Ansatz' 
\begin{eqnarray} 
\label{CEEDEEJAY}
\widetilde{\sigma}^i_a=\Psi_{ae}B^i_e.
\end{eqnarray} 
\noindent 
Equation (\ref{CEEDEEJAY}) requires that $(\hbox{det}B)$ and $(\hbox{det}\Psi)$ be nonzero, which confines the realm of its applicability to GR to spacetimes of Petrov Types I, D and O.  Substituting (\ref{CEEDEEJAY}) into (\ref{ASHVARIABLES}) and using the nondegeneracy conditions on $B$ and $\Psi$, we get the following system
\begin{eqnarray} 
\label{FINALLY}
B^i_eD_i\Psi_{ae}=0;~~\epsilon_{dae}\Psi_{ae}=0;~~\Lambda+\hbox{tr}\Psi^{-1}=0.
\end{eqnarray} 
\noindent 
In the CDJ approach, the Hamiltonian and vector constraints have been reduced to simple algebraic conditions on $\Psi$, and one is left with a set of three partial differential equations for the Gauss' law constraint.  Gauss' law has been interpreted in the literature as a condition on $\Psi$ for fixed $A$ designed to reduce $\Psi$ from five to two D.O.F., for which it is not known whether there will always exist solutions \cite{TWOFORM}, \cite{LIGHT}.  The Gauss' law constraints have remained unaddressed, and have presented a main obstacle to progress in this approach.  Moreover, as pointed out in \cite{THIEMANN} (quote) "even if one has the general solution to (\ref{FINALLY}), then one would only have obtained the constraint surface of the non-degenerate sector of the gravitational phase space, not its reduced phase space since one did not factor by the gauge orbits yet."\par 
\indent 
It is ultimately the Gauss' law constraint, along with reality conditions, within which the difficulty of the initial value constraints problem in the CDJ $A,\Psi$ variables resides.  In this paper we will obtain a reduced phase space for gravity using these variables.  We will show that there is a large solution space to all of the constraints, including the Gauss' constraint, leaving four phase space degrees of freedom per point.\par 
\noindent 
(iii) Gauge fixing: To address the observation of \cite{THIEMANN} regarding a reduced phase space we will need seven gauge fixing conditions, one condition for each constraint.  The gauge fixing conditions must be accessible, namely, it must be possible to reach the desired configuration locally from any given configuration by a suitable choice of parameters $A^a_0,N^i,N$.  In the evolution equations (\ref{USING}), spatial derivatives act on $A^a_0$ while in (\ref{THECONNECTION1}) they act on $N^i$ and $N$.  Differential operators tend to complicate the gauge-fixing problem, as they can be often nontrivial to invert.  However, let us examine the evolution equations in terms of the variables $A,\Psi$ (See Appendix A for the derivation).
The equation of motion for $A^a_i$ is given by
\begin{eqnarray} 
\label{AEQUATION}
\dot{A}^a_i=D_iA^a_0+\epsilon_{ijk}B^j_aN^k-iN\sqrt{\hbox{det}B}\sqrt{\hbox{det}\Psi}(\Psi^{-1}\Psi^{-1})^{ad}(B^{-1})^d_i,
\end{eqnarray} 
\noindent 
which is essentially identical to (\ref{USING}) in its level of difficulty (both equations have derivatives acting on $A^a_0$, with $N$ and $N^i$ free of derivatives).  The equation of motion for $\Psi$ is
\begin{eqnarray}
\label{AEQUATION1}
\dot{\Psi}_{ag}=\bigl(f_{abc}\Psi_{bg}+f_{gbc}\Psi_{ab}\bigr)A^c_0+N^jD_j\Psi_{ag}\nonumber\\
+iN\sqrt{\frac{\hbox{det}\Psi}{\hbox{det}B}}\epsilon^{fbg}(\Psi^{-1}\Psi^{-1})^{fe}B^j_bD_j\Psi_{ae}
\end{eqnarray}
\noindent 
and upon comparison, unlike in the case of (\ref{THECONNECTION1}), one sees that $N$ and $N^i$ are free of derivatives.  This means that if using other than conditions on the connection $A^a_i$ to gauge fix the vector and Hamiltonian constraints, the corresponding gauge fixing equations will be differential equations in the case of (\ref{THECONNECTION1}), but algebraic equations for (\ref{AEQUATION1}).  This general feature of algebraic versus differential equations, as we will see, is what makes the $A,\Psi$ variables ideally suited for addressing the reduced phase space problem of GR.\par 

\subsection{II.2. Transformation properties of the basic fields}

If one were to regard $A,\Psi$ as the phase space variables corresponding to a particular formulation of gravity, then the evolution equations for this formulation (\ref{AEQUATION}) and (\ref{AEQUATION1}) could be interpreted as certain transformations of the variables involving the auxiliary 
fields $\eta^a\equiv{A}^a_0$, $N^i$ and $N$
\begin{eqnarray}
\label{FINALLY1}
\dot{A}^a_i=\delta_{\vec{\eta}}A^a_i+\delta_{\vec{N}}A^a_i+\delta_NA^a_i=\delta_{Total}A^a_i;~~
\dot{\Psi}_{ae}=\delta_{\vec{\eta}}{\Psi}_{ae}+\delta_{\vec{N}}{\Psi}_{ae}+\delta_N{\Psi}_{ae}=\delta_{Total}\Psi_{ae}.
\end{eqnarray}
\noindent 
The transformations parametrized by $\eta^a$ are given by
\begin{eqnarray}
\label{NICE1}
\delta_{\vec{\eta}}A^a_i=D_i\eta^a=\partial_i\eta^a+f^{abc}A^b_i\eta^c;~~
\delta_{\vec{\eta}}\Psi_{ae}=-\eta^b\bigl(f_{abc}\Psi_{ce}+f_{ebc}\Psi_{ac}\bigr),
\end{eqnarray}
\noindent 
which for infinitesimal $\eta^a$ correspond to infinitesimal $SO(3)$ gauge transformations generated by the Gauss' law constraint.  Transformations parametrized by the shift vector $N^i$ are given by 
\begin{eqnarray} 
\label{NICE2}
\delta_{\vec{N}}A^a_i=\epsilon_{ijk}B^j_aN^k;~~
\delta_{\vec{N}}\Psi_{ae}=N^kD_k\Psi_{ae}=N^k\partial_k\Psi_{ae}+N^kA^b_k\bigl(f_{abc}\Psi_{ce}+f_{ebc}\Psi_{ac}\bigr).
\end{eqnarray}
\noindent 
For infinitesimal $N^i$, (\ref{NICE2}) correspond to infinitesimal spatial diffeomorphisms corrected by infinitesimal $SO(3)$ gauge transformations with field-dependent parameter $N^iA^b_i$, which are generated by the vector constraint.  Lastly, transformations parametrized by $N$, generated by the Hamiltonian constraint, are given by 
\begin{eqnarray} 
\label{NICE3}
\delta_NA^a_i=N(\hbox{det}B)^{1/2}\sqrt{\hbox{det}\Psi}(B^{-1})^b_i(\Psi^{-1}\Psi^{-1})^{ba}\equiv{N}H^a_i;\nonumber\\
\delta_N\Psi_{ae}=-{N}\sqrt{{{\hbox{det}\Psi} \over {\hbox{det}B}}}\epsilon^{efg}(\Psi^{-1}\Psi^{-1})^{fb}B^k_gD_k\Psi_{ab}\equiv{N}h_{ae}.
\end{eqnarray}
\noindent 
Equations (\ref{NICE1}), (\ref{NICE2}) and (\ref{NICE3}) are, strictly speaking, not gauge transformations and diffeomorphisms except in the limit where the parameters are infinitesimal.  To obtain the correct interpretation of $SO(3)$ gauge transformations and diffeomorphisms for finite parameters, the transformations must be exponentiated.  This is tantamount to integrating the (first order) evolution equations (\ref{FINALLY1}).\par 
\indent
The idea behind the $A,\Psi$ variables is that they should simplify the process of constructing solutions to the constraints (\ref{FINALLY}) and fixing a gauge through the equations (\ref{FINALLY1}), on account of the possibility of having fewer and less complicated spatial derivatives to work with.  Then one can reconstruct the spacetime metric
\begin{eqnarray} 
\label{SPACETIMEMETRIC} 
ds^2=-N^2dt^2+h_{ij}\bigl(dx^i+N^idt\bigr)\bigl(dx^j+N^jdt\bigr),
\end{eqnarray} 
\noindent 
with the spatial 3-metric given by
\begin{eqnarray} 
\label{SPACETIMEMETRIC1}
h_{ij}=(\hbox{det}\Psi)(\Psi^{-1}\Psi^{-1})^{bf}(B^{-1})^b_i(B^{-1})^f_j(\hbox{det}B).
\end{eqnarray} 
\noindent 
The lapse and shift functions in (\ref{SPACETIMEMETRIC}) fix a particular slicing of spacetime which, as we will see, is determined through the gauge-fixing equations corresponding to the gauge which will be chosen.

\section{III. Gauge-fixing considerations}

To define a reduced phase space or a sector thereof, we will satisfy the following three necessary conditions: (i) We will impose a suitable set of subsidiary (gauge-fixing) conditions on $A^a_i,\Psi_{ae}$. (ii) We will also require that the chosen gauge, which must be accessible, be preserved under time evolution. (iii) We will show that the initial value constraints admit a solution with four phase space degrees of freedom per point within this gauge.  Our choice of gauge will be motivated by the desire to gain access to nontrivial sectors of the reduced phase space for gravity, which entails constructing solutions to the constraints explicitly in terms of certain degrees of freedom.\par 
\indent  
We will require that $\Psi_{ae}$ be diagonal after gauge-fixing, namely that ${\Psi}_{(\overline{ae})}\equiv\Psi_{ae}-\delta_{ae}\lambda_e=0$, where the overline denotes the off-diagonal, symmetric part of $\Psi_{ae}$,\footnote{By the notation, the index pair $(\overline{ae})$ refers to the off-diagonal part of $\Psi_{(ae)}$, and takes on the 
values $(\overline{12})=3$, $(\overline{23})=1$ and $(\overline{31})=2$.} and $\lambda_e$ for $e=1,2,3$ are the eigenvalues.  From (\ref{NICE1}), one sees that the infinitesimal $SO(3,C)$ orbit in a neighborhood of a 
diagonal $\Psi_{ae}$ is given by $\delta_{\vec{\eta}}\Psi_{ae}\biggl\vert_{{\Psi}_{(\overline{ae})}=0}=-f_{aeb}(\lambda_a-\lambda_e)\eta^b$.  If $SO(3,C)$ transformations were the only consideration, then this transformation would be singular unless $\lambda_1\neq\lambda_2\neq\lambda_3$, which seems to naively suggest that the diagonal $\Psi$ is accessible only for Petrov Type I spacetimes.  We will return to this point later.\par 
\indent 
For gauge-fixing of the vector and Hamiltonian constraints we will choose the condition $A^a_in^i=0$ and $Q=f$, where $n^i$ is a fixed spatial 3-vector, and $Q$ is an $SO(3,C)$ invariant 
of $\Psi_{ae}$ other than $\hbox{tr}\Psi^{-1}$.  $A^a_in^i=0$ is known as the axial gauge, which occurs frequently in Yang--Mills theory.  The axial gauge requires the introduction of two external structures, namely the vector $n^i$, and an auxiliary metric $g_{ij}$ which enables the putting in place of the notion of $n^i$ as a unit vector $\vec{n}\cdot\vec{n}=g_{ij}n^in^j=1$, and $g_{ij}$ is unrelated to the 3-metric $h_{ij}$ in (\ref{SPACETIMEMETRIC1}).\par 
\indent 
From (\ref{NICE2}), one sees that $\delta_{\vec{N}}(A^a_in^i)=\epsilon_{ijk}n^iB^{ja}N^k$ is the gauge orbit of gauge-corrected spatial diffeomorphism transformations in an infinitesimal neighborhood of $A^a_in^i=0$.  This transformation cannot be solved for $\vec{n}\cdot\vec{N}=g_{ij}n^iN^j$, the component of $N^k$ in the direction of $n^i$.  Therefore $\epsilon_{ijk}n^iB^{ja}$, seen as a 3 by 3 matrix, is of rank 2.  The shift vector can be decomposed as $N^k=\overline{N}^k+(\vec{n}\cdot\vec{N})n^k$, 
where $\overline{N}^i=P^i_jN^j$, with $P^i_j=\delta^i_j-n^in_j$ as the spatial projection operator orthogonal to $n^i$.  From (\ref{NICE1}) and (\ref{NICE3}), and using $\delta_{\vec{\eta}}Q=0$ due to $SO(3,C)$ invariance, we have
\begin{eqnarray} 
\label{ORBITAL}
\delta_{N^{\mu}}(A^a_in^i)=\epsilon_{ijk}n^ib^{ja}\overline{N}^k+\delta^a_0(\vec{n}\cdot{\vec{N}})+NH^a_in^i;~~
\delta_{N^{\mu}}Q=\overline{N}^kP^i_k\partial_iQ+(\vec{n}\cdot\vec{N})n^i\partial_iQ+Nh,
\end{eqnarray} 
\noindent 
for some $h$, and a necessary condition that this transformation be nonsingular is that $n^i\partial_iQ\neq{0}$ and $h\neq{0}$.\footnote{The singular configurations form a set of measure zero, which must be avoided.}\par 
\indent 
Keeping all of the above considerations in mind, let us choose the following gauge-fixing conditions
\begin{eqnarray} 
\label{USING4} 
n^iA^a_i=0;~~\Psi_{(\overline{ae})}=0;~~Q-f=0,
\end{eqnarray} 
\noindent 
where $f$ is some chosen spacetime function with $n^i\partial_if\neq{0}$.  There are a total of seven conditions in (\ref{USING4}), one condition for each of the seven initial value constraints.  For these to be good gauge-fixing conditions, not only must (\ref{USING4}) be true, but their time derivatives must be zero so that (\ref{USING4}) are preserved under time evolution.  Taking the time derivatives and using (\ref{AEQUATION}), (\ref{AEQUATION1}) and (\ref{FINALLY1}), we have
\begin{eqnarray} 
\label{DERIVATIVE}
n^i\dot{A}^a_i=n^i\partial_i\eta^a+\epsilon_{ijk}n^iB^{ja}\overline{N}^k+\delta_a^0(\vec{n}\cdot\vec{N})+Nn^iH^a_i=0;\nonumber\\
\dot{\Psi}_{(\overline{ae})}=(-\eta^b+A^b_k\overline{N}^k)f_{aeb}(\lambda_a-\lambda_e)+Nh_{(\overline{ae})}=0~~\hbox{for}~a\neq{e};\nonumber\\
\dot{Q}-\dot{f}=\overline{N}^kP^i_k\partial_iQ+(\vec{n}\cdot\vec{N})n^i\partial_iQ+Nh=0.
\end{eqnarray} 
\noindent 
In the first term of the first line of (\ref{DERIVATIVE}) we have used $\delta_{\vec{\eta}}(A^a_in^i)\biggl\vert_{n^iA^a_i=0}=n^iD_i\eta^a=n^i\bigl(\partial_i\eta^a+f^{abc}A^b_i\eta^c\bigr)=n^i\partial_i\eta^a$, which eliminates the structure constant terms on account of the axial gauge condition.  In the second line of (\ref{DERIVATIVE}) we have used 
$\delta_{\vec{N}}{\Psi}_{(\overline{ae})}\biggl\vert_{{\Psi}_{(\overline{ae})}=0}=N^k\partial_k{\Psi}_{(\overline{ae})}
+\delta_{A^a_iN^i}{\Psi}_{(\overline{ae})}=\delta_{A^a_iN^i}{\Psi}_{(\overline{ae})}$, since $\Psi_{(\overline{ae})}=N^k\partial_k\Psi_{(\overline{ae})}=0$ on account of the gauge-fixing choice of diagonal $\Psi$.  Equation (\ref{DERIVATIVE}) can be written in the following matrix form
\begin{displaymath}
\left(\begin{array}{cccc}
\delta_{ab}n^i\partial_i & \epsilon_{ijk}n^ib^j_a & \delta_a^0 & n^iH^a_i\\
-f_{abe}(\lambda_a-\lambda_e) & f_{abe}(\lambda_e-\lambda_a)a^b_k & \delta_{(\overline{ae})}^0 &{h}_{(\overline{ae})}\\
0 & \partial_kQ & n^i\partial_iQ & h\\
\end{array}\right)
\left(\begin{array}{c}
\eta^b\\
\overline{N}^k\\
\vec{n}\cdot\vec{N}\\
N\\
\end{array}\right)
=
\left(\begin{array}{c}
\delta_a^0\\
\delta_{(\overline{ae})}^0\\
0\\
\dot{f}\\
\end{array}\right)
.
\end{displaymath}
\noindent
Irrespective of which gauge-fixing choices are associated with which particular parameters, we have arranged the gauge-fixing equations such that the differential operators occur in the matrix diagonal.  This offers the most flexibility in lifting restrictions imposed by convergence issues in inverting the matrix equation.  This is a judicious choice, as we require the set of all transformations, taken all together at once, be invertible into the desired gauge.

\subsection{III.1. Solving the gauge-fixing equations}

We will now show that the chosen gauge is accessible, namely that the gauge-fixing equations have a solution.  For purposes of transparency and without loss of generality, let us fix $n^i=(0,0,1)$.  So the gauge is $\Psi_{ae}=Diag(\lambda_1,\lambda_2,\lambda_3)$, with $Q=f$ and $A^a_i=\bigl(\delta^{ab}-\delta^{a3}\delta^{b3}\bigr)A^b_i=(A^a_1,A^a_2,\delta^a_0)$ representing a matrix of column vectors with the third column being the zero 3-vector.  For this orientation the gauge-fixing equations take the following matrix form
\begin{displaymath}
\left(\begin{array}{ccccccc}
\partial_3 & 0 & 0 & -B^2_1 & B^1_1 & 0 & H^1_3\\
0 & \partial_3 & 0 & -B^2_2 & B^1_2 & 0 & H^2_3\\
0 & 0 & \partial_3 & -B^2_3 & B^1_3 & 0 & H^3_3\\
(\lambda_3-\lambda_2) & 0 & 0 & (\lambda_2-\lambda_3)A^1_1 & (\lambda_2-\lambda_3)A^1_2 & 0 & h_{(\overline{23})}\\
0 & (\lambda_1-\lambda_3) & 0 & (\lambda_3-\lambda_1)A^2_1 & (\lambda_3-\lambda_1)A^2_2 & 0 & h_{(\overline{31})}\\
0 & 0 & (\lambda_2-\lambda_1) & (\lambda_1-\lambda_2)A^3_1 & (\lambda_1-\lambda_2)A^3_2 & 0 & h_{(\overline{12})}\\
0 & 0 & 0 & \partial_1Q & \partial_2Q & \partial_3Q & h\\
\end{array}\right)
\left(\begin{array}{c}
\eta^1\\
\eta^2\\
\eta^3\\
N^1\\
N^2\\
N^3\\
N\\
\end{array}\right)
=
\left(\begin{array}{c}
0\\
0\\
0\\
0\\
0\\
0\\
\partial{f}/\partial{t}\\
\end{array}\right)
,
\end{displaymath} 
\noindent 
where $\overline{N}^k=(N^1,N^2)$ and $\vec{n}\cdot\vec{N}=N^3$.  There are a few things to note, regarding the matrix $O$ which acts on the column 7-vector of gauge-fixing parameters. (i) The only differential operators in $O$ are directional derivative $n^i\partial_i=\partial_3\equiv\partial_z$ occurring in the upper left block 3 by 3 matrix $A\equiv\delta_{ab}\partial_z$ (along the diagonal as alluded to earlier), whose inverse is $A^{-1}=\delta_{ab}\int{dz}$ with integration constants equal to zero.\par 
\noindent 
(ii) Since the remaining sub-matrices of $O$ are algebraic, then one should expect $O$ on general grounds to be invertible.  The inverse of $O$ takes the form
\begin{displaymath}
O^{-1}=
\left(\begin{array}{cc}
A & B\\
C & D\\
\end{array}\right)
^{-1}=
\left(\begin{array}{ccc}
A^{-1}+A^{-1}BM^{-1}CA^{-1} & -A^{-1}BM^{-1}\\
-M^{-1}CA^{-1} & M^{-1}\\
\end{array}\right)
,
\end{displaymath} 
\noindent 
where $M=D-CA^{-1}B$ is the Schur complement of $A$.  A necessary and sufficient condition that $O$ be nonsingular is that $A$ and $M$ are nonsingular.\par 
\noindent 
(iii) The condition that $M$ be nonsingular reduces to the requirement that $D$, the lower right 4 by 4 block submatrix of $O$, be nonsingular.  Clearly, a necessary condition for this is that $\partial_zQ\neq{0}$.  For algebraically general spacetimes (Petrov Type I with $\lambda_1\neq\lambda_2\neq\lambda_3$), the configurations for which the minor conjugate to $\partial_zQ$ is degenerate forms a set of measure zero, which must be avoided, and similarly for Type D.  But for Type O (with $\lambda_1=\lambda_2=\lambda_3$), all possible minors of $D$ are degenerate.  So Type O spacetimes must be excluded from the set of solutions satisfying the gauge-fixing  conditions.\par 
\noindent  
(iv) Since the column 7-vector on the right hand side of the full gauge-fixing equations consists mostly of zeroes, then only the lower right block matrix of $O^{-1}$, namely $M^{-1}$, will be needed to find the lapse and shift $N,N^i$ for the purpose of constructing the spacetime metric (\ref{SPACETIMEMETRIC}).  Define the column 
4-vectors $w\equiv(N^1,N^2,N^3,N)$ and $v\equiv(0,0,0,\dot{Q})$, and let $V\rightarrow{W}$ be a linear map for normed linear spaces $V,W$, with $v\in{V}$ and $w\in{W}$.  All that is needed to show that there exists a well-defined solution for $w$ is that the map $w=M^{-1}v$ is continuous.  This is the case $if~and~only~if$ there exists an $m\in{R}$ such that $\Vert{M}^{-1}v\Vert\leq{m}\Vert{v}\Vert$ for all $v\in{V}$.  Note that the relevant part of the gauge fixing equation satisfies the norm inequality
\begin{eqnarray} 
\label{NORMINEQUALITY}
\Vert{w}\Vert\leq\Vert{M}^{-1}\Vert\Vert{v}\Vert,
\end{eqnarray}
\noindent
where $\Vert{u}\Vert$ is an appropriately defined norm for the respective vector space.  Then one needs only choose $m=\Vert{M}^{-1}\Vert$ and one is done.\par 
\indent 
We must now define the norm $\Vert{M}^{-1}\Vert$, and show that it is finite.  We have already argued that $A$ is invertible, since $A^{-1}\equiv{I}\int{dz}$ with $I$ the identity operator, and now we must show that $M$ is also invertible.  Note that the action of $M^{-1}$ can be written formally as an operator geometric series
\begin{eqnarray} 
\label{COMPLEMENT}
M^{-1}v=\bigl(1-D^{-1}CA^{-1}B\bigr)^{-1}D^{-1}v=\sum_{n=0}^{\infty}\bigl(D^{-1}CA^{-1}B)^nD^{-1}v.
\end{eqnarray}
\noindent 
For this to be well-defined we must require that $\hbox{det}D\neq{0}$, so that $D^{-1}$ exists, which seems to be generically true (except on a set of measure zero, which is to be avoided).  By the extreme value theorem, any continuous function on a closed region $S\in\Sigma$ must be bounded above and below.  Since we require our GR solutions to consist of well-defined, continuous functions (we will further restrict them to be real-analytic due to the Gauss' law constraint equations), then we can 
define $\alpha\equiv\Vert{D}^{-1}\Vert\Vert{C}\Vert\Vert{B}\Vert$, with the norm of any matrix $Q$ defined as 
\begin{eqnarray} 
\label{NORMS}
\Vert{Q}\Vert=sup\{\sum_{a,b}\vert{Q}_{ab}(x,y,z)\vert\}_{\forall{x,y,z\in{S}}}.
\end{eqnarray}
\noindent 
Using (\ref{NORMS}), then (\ref{COMPLEMENT}) satisfies the following norm inequality
\begin{eqnarray} 
\label{COMPLEMENTONE}
\Vert{M}^{-1}v\Vert\leq\sum_{n=0}^{\infty}(\alpha\int_0^{\vert{z}\vert}{dz}^{\prime})^n\Vert{D}\Vert^{-1}\Vert{v}\Vert=e^{\alpha\vert{z}\vert}\Vert{D}^{-1}\Vert\Vert{v}\Vert,
\end{eqnarray}
\noindent
where $(\int{dz})^n\equiv\int{dz}\int{dz}\dots\int{dz}$.  So a sufficient condition for the ability to gauge-fix as indicated, reading off the lapse and shift vector, is that
\begin{eqnarray}
\label{COMPLEMENTTWO}
m=e^{\alpha\vert{z}\vert}\Vert{D}^{-1}\Vert<\infty.
\end{eqnarray} 
\noindent 
Equation (\ref{COMPLEMENTTWO}) is certainly satisfied for compact spatial 3-manifolds $\Sigma$, which means that (\ref{COMPLEMENT}) is well-defined, and also provides a way to find the lapse and shift.  For the noncompact case, a tighter bound would be needed.

\section{IV. The initial value constraints}

The initial value constraints in $A,\Psi$ variables can be written entirely in the language of the 3-dimensional special complex orthogonal group $SO(3,C)$, which contains no direct reference to a metric or to coordinates.  These are given by (\ref{FINALLY1}), re-written here for completeness
\begin{eqnarray}
\label{VALUE}
B^i_eD_i\Psi_{ae}=0;~~\epsilon_{dae}\Psi_{ae}=0;~~\Lambda+\hbox{tr}\Psi^{-1}=0.
\end{eqnarray}
\noindent
The Hamiltonian and vector constraints are simple algebraic equations, which are straightforward to solve.  This leaves remaining the Gauss' law constraint as the system of partial differential equations
\begin{eqnarray}
\label{GOOSE}
B^i_eD_i\Psi_{ae}=B^i_e\partial_i\Psi_{ae}+C_{be}\bigl(f_{abf}\Psi_{fe}+f_{ebg}\Psi_{ag}\bigr)=0,
\end{eqnarray}
\noindent 
where we have defined the helicity density matrix $C_{ae}$ by
\begin{eqnarray}
\label{VALUE1}
C_{ae}=A^a_iB^i_e=\epsilon^{ijk}A^a_i\partial_jA^e_k+\delta_{ae}(\hbox{det}A).
\end{eqnarray}
\noindent 
The vector constraint states that $\Psi_{ae}=\Psi_{ea}$ must be symmetric.  Using a triple of complex rotation parameters $\theta^a$, a complex symmetric matrix can be written as a polar decomposition
\begin{eqnarray} 
\label{DECOMP}
\Psi_{ae}=(e^{\theta\cdot{T}})_{af}\lambda_f(e^{-\theta\cdot{T}})_{fe},
\end{eqnarray} 
\noindent 
where $T_a$ are the $SO(3)$ generators and $\lambda_f$ are the eigenvalues of $\Psi_{ae}$.  Due to the cyclic property of the trace, $\theta^a$ cancels out of the Hamiltonian contraint, yielding the following algebraic relation amongst the eigenvalues
\begin{eqnarray} 
\label{DECOMP1}
\Lambda+{1 \over {\lambda_1}}+{1 \over {\lambda_2}}+{1 \over {\lambda_3}}=0.
\end{eqnarray} 
\noindent

\subsection{IV.1. Solving the Gauss' law constraint}

We have shown that the gauge $A^a_in^i=A^a_3=0$ with $\Psi_{ae}=Diag(\lambda_1,\lambda_2,\lambda_3)$ is accessible and is preserved under time evolution.  We will next show that the initial value constraints have a solution within this gauge.  Since $\Psi_{ae}=\delta_{ae}\lambda_e$ is diagonal, then the Gauss' law constraint reduces to
\begin{eqnarray} 
\label{GOOSEIT}
B^i_1\partial_i\lambda_1+(\lambda_3-\lambda_1)C_{23}+(\lambda_1-\lambda_2)C_{32}=0;\nonumber\\
B^i_2\partial_i\lambda_2+(\lambda_1-\lambda_2)C_{31}+(\lambda_2-\lambda_3)C_{13}=0;\nonumber\\
B^i_3\partial_i\lambda_3+(\lambda_2-\lambda_3)C_{12}+(\lambda_3-\lambda_1)C_{21}=0,
\end{eqnarray}
\noindent 
where the helicity density matrix elements in this gauge are given by
\begin{eqnarray} 
\label{GOOSEIT1} 
C_{12}=-A^1_1\partial_3A^2_2+A^1_2\partial_3A^2_1;~~C_{21}=-A^2_1\partial_3A^1_2+A^2_2\partial_3A^1_1;\nonumber\\
C_{23}=-A^2_1\partial_3A^3_2+A^2_2\partial_3A^3_1;~~C_{32}=-A^3_1\partial_3A^2_2+A^3_2\partial_3A^2_1;\nonumber\\
C_{31}=-A^3_1\partial_3A^1_2+A^3_2\partial_3A^1_1;~~C_{13}=-A^1_1\partial_3A^3_2+A^1_2\partial_3A^3_1.
\end{eqnarray}
\noindent 
Substituting (\ref{GOOSEIT1}) into (\ref{GOOSEIT}), then the Gauss' law constraint becomes vastly simplified.  Let us make the identifications $x^1\equiv{x}$, $x^2\equiv{y}$, and $x^3\equiv{z}$ and expand it out.  Then the first Gauss' law constraint equation is
\begin{eqnarray} 
\label{GAUSSES1}
\Bigl(\frac{\partial\lambda_1}{\partial{y}}\frac{\partial{A}^1_1}{\partial{z}}-\frac{\partial{A}^1_1}{\partial{y}}\frac{\partial\lambda_1}{\partial{z}}\Bigr)
+\Bigl(\frac{\partial\lambda_1}{\partial{z}}\frac{\partial{A}^1_2}{\partial{x}}-\frac{\partial{A}^1_2}{\partial{z}}\frac{\partial\lambda_1}{\partial{x}}\Bigr)\nonumber\\
+\bigl(A^2_1A^3_2-A^2_2A^3_1\bigr)\frac{\partial\lambda_1}{\partial{z}} 
+(\lambda_3-\lambda_1)\Bigl(-A^2_1\frac{\partial{A}^3_2}{\partial{z}}+A^2_2\frac{\partial{A}^3_1}{\partial{z}}\Bigr)+(\lambda_1-\lambda_2)\Bigl(-A^3_1\frac{\partial{A}^2_2}{\partial{z}}+A^3_2\frac{\partial{A}^2_1}{\partial{z}}\Bigr)=0.
\end{eqnarray}
\noindent 
The second Gauss' law constraint equation is
\begin{eqnarray} 
\label{GAUSSES11}
\Bigl(\frac{\partial\lambda_2}{\partial{y}}\frac{\partial{A}^2_1}{\partial{z}}-\frac{\partial{A}^2_1}{\partial{y}}\frac{\partial\lambda_2}{\partial{z}}\Bigr)
+\Bigl(\frac{\partial\lambda_2}{\partial{z}}\frac{\partial{A}^2_2}{\partial{x}}-\frac{\partial{A}^2_2}{\partial{z}}\frac{\partial\lambda_2}{\partial{x}}\Bigr)\nonumber\\
+\bigl(A^3_1A^1_2-A^3_2A^1_1\bigr)\frac{\partial\lambda_2}{\partial{z}} 
+(\lambda_1-\lambda_2)\Bigl(-A^3_1\frac{\partial{A}^1_2}{\partial{z}}+A^3_2\frac{\partial{A}^1_1}{\partial{z}}\Bigr)+(\lambda_2-\lambda_3)\Bigl(-A^1_1\frac{\partial{A}^3_2}{\partial{z}}+A^1_2\frac{\partial{A}^3_1}{\partial{z}}\Bigr)=0,
\end{eqnarray}
\noindent 
and the third Gauss' law constraint equation is
\begin{eqnarray} 
\label{GAUSSES12}
\Bigl(\frac{\partial\lambda_3}{\partial{y}}\frac{\partial{A}^3_1}{\partial{z}}-\frac{\partial{A}^3_1}{\partial{y}}\frac{\partial\lambda_3}{\partial{z}}\Bigr)
+\Bigl(\frac{\partial\lambda_3}{\partial{z}}\frac{\partial{A}^3_2}{\partial{x}}-\frac{\partial{A}^3_2}{\partial{z}}\frac{\partial\lambda_3}{\partial{x}}\Bigr)\nonumber\\
+\bigl(A^1_1A^2_2-A^1_2A^2_1\bigr)\frac{\partial\lambda_3}{\partial{z}} 
+(\lambda_2-\lambda_3)\Bigl(-A^1_1\frac{\partial{A}^2_2}{\partial{z}}+A^1_2\frac{\partial{A}^2_1}{\partial{z}}\Bigr)+(\lambda_3-\lambda_1)\Bigl(-A^2_1\frac{\partial{A}^1_2}{\partial{z}}+A^2_2\frac{\partial{A}^1_1}{\partial{z}}\Bigr)=0.
\end{eqnarray}
\noindent
It was mentioned earlier that the existing references \cite{TWOFORM}, \cite{LIGHT} regard Gauss' law as a set of differential equations for three components of $\Psi_{ae}$ for fixed $A^a_i$, for which it is not known whether a general solution exists.  The result of the present paper, which takes gauge-fixing into account, implies instead the following way to interpret Gauss' law: Fix the eigenvalues $\lambda_1,\lambda_2,\lambda_3$ via the Hamiltonian constraint (\ref{DECOMP1}) combined with the gauge-fixing condition $Q=f$ and fix three elements, $u$ of $A^a_i$.  Then Gauss' law is a set of linear, first order PDEs to be solved for the remaining three components $v$ of the connection $A^a_i$ in terms of them.  To see this, it is instructive to rearrange Gauss' law (\ref{GAUSSES1}), (\ref{GAUSSES11}) and (\ref{GAUSSES12}) into the following matrix form.  Define the following matrices 

\begin{displaymath}
A^{\prime}=
\left(\begin{array}{ccc}
\partial_y\lambda_1 & (\lambda_1-\lambda_2)A^3_2 & (\lambda_3-\lambda_1)A^2_2\\
(\lambda_1-\lambda_2)A^3_2 & \partial_y\lambda_2 & (\lambda_2-\lambda_3)A^1_2\\
(\lambda_3-\lambda_1)A^2_2 & (\lambda_2-\lambda_3)A^1_2 & \partial_y\lambda_3\\
\end{array}\right)
;~~C^{\prime}=
\left(\begin{array}{ccc}
-\partial_z\lambda_1 & 0 & 0\\
0 & -\partial_z\lambda_2 & 0\\
0 & 0 & -\partial_z\lambda_3\\
\end{array}\right)
,
\end{displaymath}

\begin{displaymath}
B^{\prime}=
\left(\begin{array}{ccc}
0 & A^3_2\partial_z\lambda_1-(\lambda_3-\lambda_1)\partial_zA^3_2 & -\bigl(A^2_2\partial_z\lambda_1+(\lambda_1-\lambda_2)\partial_zA^2_2\bigr)\\
-\bigl(A^3_2\partial_z\lambda_2+(\lambda_2-\lambda_3)\partial_zA^3_2\bigr) & 0 & A^1_2\partial_z\lambda_2-(\lambda_1-\lambda_2)\partial_zA^1_2\\
A^2_2\partial_z\lambda_3-(\lambda_2-\lambda_3)\partial_zA^2_2 & -\bigl(A^1_2\partial_z\lambda_1+(\lambda_3-\lambda_1)\partial_zA^1_2\bigr) & 0\\
\end{array}\right)
,
\end{displaymath} 

\noindent 
and the following column 3-vectors

\begin{displaymath}
v=
\left(\begin{array}{ccc}
A^1_1\\
A^2_1\\
A^3_1\\
\end{array}\right)
;~~u=
\left(\begin{array}{ccc}
A^1_2\\
A^2_2\\
A^3_2\\
\end{array}\right)
;~~D^{\prime}=
\left(\begin{array}{c}
(\partial_z\lambda_1)\partial_xA^1_2-(\partial_x\lambda_1)\partial_zA^1_2\\
(\partial_z\lambda_2)\partial_xA^2_2-(\partial_x\lambda_2)\partial_zA^2_2\\
(\partial_z\lambda_3)\partial_xA^3_2-(\partial_x\lambda_3)\partial_zA^3_2\\
\end{array}\right)
.
\end{displaymath} 
\noindent 
We will see that this way of interpreting the Gauss' law constraint guarantees under fairly general considerations the existence of solutions featuring four free functions per spatial point, and we will regard these as the physical degrees of freedom for the reduced phase space of GR in this gauge.  Utilizing the above definitions, then the Gauss' law constraint can be written as the following system
\begin{eqnarray} 
\label{SYSTEM}
A^{\prime}\partial_zv+B^{\prime}v+c^{\prime}\partial_yv+D^{\prime}=0.
\end{eqnarray}
\noindent 
For $\hbox{det}A^{\prime}$ nonzero, then $A'$ is invertible and we can construct ${A'}^{-1}$ and make the following definitions\footnote{The configurations for which $\hbox{det}A'=0$ form a set of measure zero, which must be avoided.}
\begin{eqnarray} 
\label{SYSTEM1}
A\equiv-{A'}^{-1}C'\partial_y-{A'}^{-1}B';~~B\equiv-{A'}^{-1}D'.
\end{eqnarray}
\noindent
Then the Gauss' law constraint can further be written in the form
\begin{eqnarray} 
\label{SYSTEM2}
\partial_zv=Av+B.
\end{eqnarray} 
\noindent 
Given one arbitrary real-analytic function $\lambda_1(x,y,z)$ corresponding to one eigenvalue of $\Psi_{ae}$. then this determines $\lambda_2$ and $\lambda_3$ uniquely via the Hamiltonian constraint (\ref{DECOMP1}) and its gauge-fixing condition $Q=f,~\partial_zf\neq{0}$.  Substituting these values into (\ref{GAUSSES1}), (\ref{GAUSSES11}) and (\ref{GAUSSES12}) along with three real-analytic connection components $u^a(x,y,z)=A^a_2(x,y,z)$, and given analytic boundary data $v^a(x,y,0)\equiv{A}^a_1(x,y,0)$ on the x-y plane (namely the 2-dimensional hypersurface orthogonal to the gauge-fixing direction $n^i$, along with the differential equation (\ref{SYSTEM2}), then by the Cauchy--Kovalevskaya theorem \cite{CK}, there exists within some neighborhood of $z=0$ a solution to the system of differential equations (\ref{SYSTEM2}).  Moreover, the solution is unique and analytic.\footnote{Note that this requires that $\lambda_2$ and $\lambda_3$, as determined from (\ref{DECOMP1}) once $\lambda_1$ is chosen, be analytic.  This can be phrased as an algebraic restriction on the allowed values of $\lambda_1$ (See. e.g. (\ref{CANBEWRITTEN2}).}\par 
\indent 
So we have a solution, in some region of 3-space $\Sigma$, a solution to the initial value constraints within the chosen gauge featuring four free analytic functions featuring four phase space degrees of freedom per spatial point.  To construct the solution one may evaluate consecutive partial derivatives of (\ref{SYSTEM2}) with respect to $z$, iterating the right hand side, and construct a Taylor expansion in $z$.  The input from the physical degrees of freedom is inherent in the operator $A=A[u,\lambda]$ and the matrix $B[u,\lambda]$, where $\lambda$ represent the eigenvalues of $\Psi_{ae}$, and $v$ can be expressed completely in terms of these quantities and the boundary data, and their spatial derivatives.\par 
\indent 
The solution for $v$ can be written as a Taylor series
\begin{eqnarray} 
\label{SYSTEM3}
v(x,y,z)=\sum_{m,n\geq{0}}^{\infty}v_{mn}(x)y^mz^n,
\end{eqnarray}
\noindent 
with 
\begin{eqnarray} 
\label{SYSTEM4}
v_{mn}=\frac{1}{m!n!}\frac{\partial^{m+n}v}{\partial{y}^m\partial{z}^n}\biggl\vert_{y=z=0}.
\end{eqnarray}
\noindent 
All partial $z$ derivatives can be obtained via recursion.  For example, for the second partial one acts on (\ref{SYSTEM2}) to obtain
\begin{eqnarray} 
\label{SYSTEM5}
\frac{\partial^2v}{\partial{z}^2}=\frac{\partial{A}}{\partial{z}}v+A\frac{\partial{v}}{\partial{z}}+\frac{\partial{B}}{\partial{z}}\nonumber\\
=\frac{\partial{A}}{\partial{z}}v+A(Av+B)+\frac{\partial{B}}{\partial{z}}=\Bigl(\frac{\partial{A}}{\partial{z}}+A^2\Bigr)v+\frac{\partial{B}}{\partial{z}}+AB,
\end{eqnarray}
\noindent 
evaluated at $z=0$.  One can continue the iteration process to any order desired, and each operation is well-defined and explicitly in terms of the boundary data.\par 
\indent
A compact way of writing down the solution to (\ref{SYSTEM2}) is in integral form
\begin{eqnarray} 
\label{INTEGRAL}
v^a(x,y,z)=v^a(x,y,0)+\int^z_0dz'B^a(x,y,z')+\int^z_0dz'A^a_b(x,y,z')v^b(x,y,z'),
\end{eqnarray}
\noindent 
where $A$ is a matrix differential operator proportional to $\partial_y$.  Iterating (\ref{INTEGRAL}), one can define a z-ordered propagator
\begin{eqnarray} 
\label{INTEGRAL1}
U^a_b(x,y;z,0)=Z\hbox{exp}\Bigl[\int^z_0dz'A(x,y,z')\Bigr],
\end{eqnarray} 
\noindent 
and the Gauss' law can formally be written in the following compact way
\begin{eqnarray} 
\label{INTEGRAL2}
v^a(x,y,z)=U^a_b(x,y;z,0)\Bigl[v^b(x,y,0)+\int^z_0dz'B^b(x,y,z')\Bigr].
\end{eqnarray} 
\noindent 
The functions $v^a(x,y,0)$ represent real-analytic boundary data on the $z=0$ hyperplane, which is freely specifiable.  Equation (\ref{INTEGRAL2}), while a formal expression, in this application is well-defined, and could provide a systematic way of organizing the terms if one wishes to attempt constructing solutions in practice.  

\section{V. Discussion and Summary}

The results of this paper have demonstrated that the initial value constraints problem of four dimensional, Lorentzian signature GR is within reach.  The main outstanding problem has been the construction of solutions to the Gauss' law constraint, which in the CDJ $A,\Psi$ variables consists of a system of three (seemingly nonlinear) partial differential equations for the components of the Ashtekar connection $A^a_i$ in terms of the eigenvalues of $\Psi_{ae}$.  We have found a particular gauge choice, the axial gauge, which reduces the Gauss' law constraint to a linear system for three components of $A^a_i$ with all other quantities freely specifiable, subject only to the requirements of (i) being real-analytic functions, (ii) various configurations, of measure zero, to be avoided guaranteeing the nondegeneracy of various parts of the gauge-fixing equations.  These include configurations such as

\begin{displaymath}
\hbox{det}
\left(\begin{array}{ccc}
(\lambda_2-\lambda_3)A^1_1 & (\lambda_2-\lambda_3)A^1_2 & h_{(\overline{23})}\\
(\lambda_3-\lambda_1)A^2_1 & (\lambda_3-\lambda_1)A^2_2 & h_{(\overline{31})}\\
(\lambda_1-\lambda_2)A^3_1 & (\lambda_1-\lambda_2)A^3_2 & h_{(\overline{12})}\\
\end{array}\right)
=0,
\end{displaymath} 
\noindent 
which would render the gauge-fixing equations singular, 

\begin{displaymath}
\hbox{det}
\left(\begin{array}{ccc}
\partial_y\lambda_1 & (\lambda_1-\lambda_2)A^3_2 & (\lambda_3-\lambda_1)A^2_2\\
(\lambda_1-\lambda_2)A^3_2 & \partial_y\lambda_2 & (\lambda_2-\lambda_3)A^1_2\\
(\lambda_3-\lambda_1)A^2_2 & (\lambda_2-\lambda_3)A^1_2 & \partial_y\lambda_3\\
\end{array}\right)
=0,
\end{displaymath} 
\noindent 
which would invalidate the conditions required for the Cauchy--Kovalevskaya theorem.  Avoidance of these configurations must be checked for each individual solution to the constraints.\par 
\indent 
Additionally, the requirement of analyticity of the eigenvalues $\lambda_f$ should be checked for the chosen function $Q$.  For example, let us choose $Q=\hbox{det}\Psi=\lambda_1\lambda_2\lambda_3$ with $\partial_zQ\neq{0}$.  Using the characteristic equation for 3 by 3 matrices, the Hamiltonian constraint can be written as
\begin{eqnarray} 
\label{CANBEWRITTEN}
\bigl(\hbox{tr}\Psi^2-(\hbox{tr}\Psi)^2\bigr)+2\Lambda(\hbox{det}\Psi)=0.
\end{eqnarray}
\noindent
Using the gauge condition $Q=f$, we have
\begin{eqnarray} 
\label{CANBEWRITTEN1}
\lambda_1\lambda_2+\lambda_2\lambda_3+\lambda_3\lambda_1+2\Lambda{Q}=\lambda_1\lambda_2+\frac{f}{\lambda_1}+\frac{f}{\lambda_2}+2\Lambda{f}=0.
\end{eqnarray} 
\noindent 
This implies a quadratic equation for $\lambda_2=\lambda_2(\lambda_1,f)$, with solution
\begin{eqnarray} 
\label{CANBEWRITTEN2}
\lambda_2=\frac{-f\Bigl(\frac{1}{\lambda_1}+2\Lambda\Bigr)\pm\sqrt{f^2\Bigl(\frac{1}{\lambda_1}+2\Lambda\Bigr)^2-4\lambda_2f}}{2\lambda_1}.
\end{eqnarray}
\noindent 
In meeting the requirements of the Cauchy--Kovalevskaya theorem for Gauss' law we must require $f$ and $\lambda_1$ to be real analytic functions.  Then the existence of a solution requires for the 
choice $\lambda_1,f$ that $\lambda_2$ as determined by (\ref{CANBEWRITTEN2}) and $\lambda_3=f/\lambda_1\lambda_2$ also be analytic.  While in the general case there will be radicals involved, they can in-principle be expanded in Taylor series with some nonzero radius of convergence.\par 
\noindent 
(iii) We have also restricted consideration to the time symmetric case, in the interest of real solutions of Lorentzian signature.  This includes initial data which is the evolution of time symmetric data.\par \noindent 
(iv) Finally, our results do not apply to other than spacetimes of Petrov Types I, D and O, due to nondegeneracy conditions required for the CDJ variables.\par 
\indent 

\section{Appendix A. Derivation of the $A,\Psi$ evolution equations}

Starting from the Lagrange equations for the Ashtekar variables (\ref{USING}) and (\ref{THECONNECTION1}), repeated here for completeness
\begin{eqnarray}
\label{DERIVATION}
\dot{A}^a_i=D_iA^a_0+\epsilon_{ijk}B^j_aN^k-iN\sqrt{\hbox{det}\widetilde{\sigma}}B^n_d(\widetilde{\sigma}^{-1})^d_i(\widetilde{\sigma}^{-1})^a_n,
\end{eqnarray} 
\noindent 
and 
\begin{eqnarray} 
\label{DERIVATION1}
\dot{\widetilde{\sigma}^i_a}=f_{abc}\widetilde{\sigma}^i_bA^c_0+\epsilon_{mjk}\epsilon^{jni}D_n(\widetilde{\sigma}^m_aN^k)
-i\epsilon^{ijk}D_j(N\sqrt{\hbox{det}\widetilde{\sigma}}(\widetilde{\sigma}^{-1})^a_k),
\end{eqnarray}
\noindent 
we will derive the evolution equations for the variables $A$ and $\Psi$.  First, recall the Ansatz (\ref{CEEDEEJAY}), also repeated here for completeness
\begin{eqnarray} 
\label{DERIVATION2}
\widetilde{\sigma}^i_a=\Psi_{ae}B^i_e.
\end{eqnarray}
\noindent 
Substitution of (\ref{DERIVATION2}) into (\ref{DERIVATION}) yields the following equation for $A^a_i$ on the phase space $A,\Psi$
\begin{eqnarray} 
\label{DERIVATION3}
\dot{A}^a_i=D_iA^a_0+\epsilon_{ijk}B^j_aN^k-iN\sqrt{\hbox{det}B}\sqrt{\hbox{det}\Psi}(\Psi^{-1}\Psi^{-1})^{ad}(B^{-1})^d_i  
\end{eqnarray} 
\noindent 
Now we will derive the evolution equation for $\Psi$.  Applying the Leibniz rule, the time derivative of (\ref{DERIVATION2}) is given by
\begin{eqnarray} 
\label{DERIVATION4}
\dot{\widetilde{\sigma}^i_a}=\dot{\Psi}_{ae}B^i_e+\Psi_{ae}\dot{B}^i_e=\dot{\Psi}_{ae}B^i_e+\Psi_{ae}\epsilon^{ijk}D_j\dot{A}^e_k.
\end{eqnarray}
\noindent 
Substituting (\ref{DERIVATION}) into (\ref{DERIVATION4}) we have\footnote{We will substitute (\ref{DERIVATION2}) near the end of the derivation, for convenience.}
\begin{eqnarray} 
\label{DERIVATION5}
\dot{\widetilde{\sigma}^i_a}=\dot{\Psi}_{ae}B^i_e+\Psi_{ae}\epsilon^{ijk}D_j\Bigl(D_kA^e_0+\epsilon_{kmn}B^m_eN^n-iN\sqrt{\hbox{det}\widetilde{\sigma}}
\Psi_{ae}B^n_d(\widetilde{\sigma}^{-1})^d_k(\widetilde{\sigma}^{-1})^e_n\Bigr)\nonumber\\
=\dot{\Psi}_{ae}B^i_e+\Psi_{ae}f^{ebg}B^i_bA^g_0
+\Psi_{ae}\epsilon^{ijk}D_j\Bigl(\epsilon_{kmn}B^m_eN^n-iN\sqrt{\hbox{det}\widetilde{\sigma}}
\Psi_{ae}B^n_d(\widetilde{\sigma}^{-1})^d_k(\widetilde{\sigma}^{-1})^e_n\Bigr)
\end{eqnarray}
\noindent 
where we have used $\epsilon^{ijk}D_jD_kA^e_0=f^{ebg}B^i_bA^g_0$, namely the definition of curvature as the commutator of two covariant derivatives.\par 
\indent 
Next we will apply the Leibniz rule to the last terms of (\ref{DERIVATION5}), bringing $\Psi_{ae}$ into the large brackets and subtracting the remainder.  This yields
\begin{eqnarray} 
\label{DERIVATION6}
\dot{\widetilde{\sigma}^i_a}=\dot{\Psi}_{ae}B^i_e+\Psi_{ae}f^{ebg}B^i_bA^g_0\nonumber\\
+D_j\Bigl(\epsilon^{ijk}\epsilon_{kmn}\Psi_{ae}B^m_eN^n-i\epsilon^{ijk}N\sqrt{\hbox{det}\widetilde{\sigma}}\Psi_{ae}B^n_d(\widetilde{\sigma}^{-1})^d_k(\widetilde{\sigma}^{-1})^e_n\Bigr)\nonumber\\
-\Bigl(\epsilon^{ijk}\epsilon_{kmn}B^m_eN^n-i\epsilon^{ijk}N\sqrt{\hbox{det}\widetilde{\sigma}}B^n_d(\widetilde{\sigma}^{-1})^d_k(\widetilde{\sigma}^{-1})^e_n\Bigr)D_j\Psi_{ae}.
\end{eqnarray}
\noindent 
So we have two expressions for $\dot{\widetilde{\sigma}^i_a}$, namely (\ref{DERIVATION1}) and (\ref{DERIVATION6}), which we can set equal to each other.  Using (\ref{DERIVATION2}) in the 
middle line of (\ref{DERIVATION6}) and in the first term of (\ref{DERIVATION1}) and upon relabelling of indices, one sees that the second and third terms on the right hand side of (\ref{DERIVATION1}) are the same as the middle line of (\ref{DERIVATION6}).  So cancelling these terms out, we are left with the following relation
\begin{eqnarray} 
\label{DERIVATION7}
f_{abc}\Psi_{be}B^i_eA^c_0=\dot{\Psi}_{ae}B^i_e+\Psi_{ae}f^{ebc}B^i_bA^c_0\nonumber\\
-\bigl(\delta^i_m\delta^j_n-\delta^i_n\delta^j_m\bigr)B^m_eN^nD_j\Psi_{ae}+iN\sqrt{\hbox{det}\widetilde{\sigma}}\epsilon^{ijk}(\Psi^{-1})^{ed}(\widetilde{\sigma}^{-1})^d_kD_j\Psi_{ae}.
\end{eqnarray}
\noindent 
Multiplying (\ref{DERIVATION7}) by $(B^{-1})^g_i$ and rearranging, we get the following evolution equation for $\Psi_{ag}$
\begin{eqnarray} 
\label{DERIVATION8}
\dot{\Psi}_{ag}=\bigl(f_{abc}\Psi_{bg}+f_{gbc}\Psi_{ab}\bigr)A^c_0+N^jD_j\Psi_{ag}-N^i(B^{-1})^g_iB^j_eD_j\Psi_{ae}\nonumber\\
+iN\sqrt{\frac{\hbox{det}\Psi}{\hbox{det}B}}\epsilon^{fbg}(\Psi^{-1}\Psi^{-1})^{fe}B^j_bD_j\Psi_{ae}.
\end{eqnarray}

\section{Acknowledgements}
This work has been supported in part by the Office of Naval Research
under Grant No. N-00-1613-WX-20992.  The author is also grateful to
Kirill Krasnov for insightful discussions and well as for helpful guidance, 
comments and criticisms on the manuscript.


\begin{thebibliography}{99}

\bibitem{DIR} Dirac P 1964 {\it Lectures on quantum mechanics} (Yeshiva University Press, New York)

\bibitem{ASH1} {Abhay Ashtekar. `New perspectives in canonical gravity', (Bibliopolis, Napoli, 1988).}

\bibitem{ASH2} {Abhay Ashtekar `New Hamiltonian formulation of general relativity'
Phys. Rev. D36 (1987) 1587}

\bibitem{ASH3} {Abhay Ashtekar `New variables for classical and quantum gravity'
Physical Review Letters 57, 18 (1986)}

\bibitem{TWOFORM} {Riccardo Capovilla, John Dell and Ted Jacobson `Self-dual 2-forms and gravity'
Class. Quantum Grav. 8 (1991) 41-57}

\bibitem{LIGHT} Capovilla R, Dell J and Jacobson T The initial value problem in light of Ashtekar's variables' {\it Preprint} gr-qc/9302020

\bibitem{THIEMANN} {Thomas Thiemann `On the solution of the initial value constraints for general relativity coupled to matter in terms of Ashtekar variables'
Class. Quantum Grav. 10: 1907-1921 (1993)}

\bibitem{CK} {E.C. Zachmanoglou and Dale W. Thoe `Introduction to Partial Differential Equations with Applications'
Dover Publication, Inc., New York (1986)}




\end{thebibliography}
\end{document}